\documentclass[reprint, 10pt, a4paper, aip, floatfix, longbibliography]{revtex4-2}
\usepackage[utf8]{inputenc}
\usepackage{amsmath}
\usepackage{amsfonts}
\usepackage{amssymb}
\usepackage{mathrsfs}
\usepackage{subfigure}
\usepackage[hidelinks]{hyperref}
\usepackage{outlines}
\usepackage{tikz}
\usepackage{booktabs}

\newcommand{\bx}{\mathbf{x}}

\newcommand{\bv}{\mathbf{v}}
\newcommand{\bu}{\mathbf{u}}

\newcommand{\D}{\mathrm{d}}

\begin{document}

\title{Upper and Lower Bounds on Phase-Space Rearrangements}
\date{\today}
\author{E.~J. Kolmes}
\email[Electronic mail: ]{ekolmes@princeton.edu}
\affiliation{Department of Astrophysical Sciences, Princeton University, Princeton, New Jersey 08544, USA}
\author{N.~J.~Fisch}
\affiliation{Department of Astrophysical Sciences, Princeton University, Princeton, New Jersey 08544, USA}

\begin{abstract}
Broad classes of plasma phenomena can be understood in terms of phase-space rearrangements.
For example, the net effect of a wave-particle interaction may consist of moving populations of particles from one region of phase space to another. 
Different phenomena drive rearrangements that obey different rules. 
When those rules can be specified, it is possible to calculate bounds that limit the possible effects the rearrangement could have (such as limits on how much energy can be extracted from the particles). 
This leads to two problems. 
The first is to understand the mapping between the allowed class of rearrangements and the possible outcomes that these rearrangements can have on the overall distribution. 
The second is to understand which rules are appropriate for which physical systems. 
There has been recent progress on both fronts, but a variety of interesting questions remain. 
\end{abstract}

\maketitle

\section{Introduction}

Consider the interaction of a wave with a plasma. 
Depending on the details of the interaction, there can be a transfer of energy between the plasma and the wave, with the wave either being amplified or damped. 
If the wave is amplified, the maximum amplification is set by how much energy it can remove from the plasma. 
If, in particular, the wave is fed by the kinetic energy of the plasma particles, this means that the wave-particle interaction somehow rearranges those particles in phase space so as to liberate some of their energy. 

Given basic rules for how that phase-space rearrangement takes place, it is possible to calculate bounds on what effects those rearrangements can have. 
For example, this can lead to limits on how much energy can be extracted from a given distribution of particles for very general classes of wave-particle interactions. 
This allows us to formalize and quantify the idea that some distributions have more accessible energy than others (for example, a bump-on-tail distribution compared with a Maxwellian). 

There are different rules that we could pick for these rearrangements, appropriate for different situations, and the ways in which different rules change the possible outcomes is often nontrivial. 
One simple rule, sometimes called Gardner restacking, is to permit any rearrangement of phase space that respects Liouville's theorem.\cite{Gardner1963} 
Another, sometimes called diffusive exchange, models phase-mixing processes by instead averaging the contents of phase-space elements.\cite{Fisch1993} 
Either basic rule can be modified by the imposition of further constraints such as conservation laws\cite{Helander2017ii, Helander2020, Kolmes2020ConstrainedDiffusion, Kolmes2024Flutes} (for example, requiring that any rearrangements must preserve one or more adiabatic invariants). 
Different authors use different conventions, but the energy that can be extracted is sometimes called the free or available energy. 
We will use these terms interchangeably here. 
The energy that can be extracted using Gardner's restacking operations is sometimes called the Gardner free or available energy; the energy that can be extracted using mixing operations is sometimes called the diffusive free or available energy. 

This leads to two essential problems. 
The first is to understand the range of outcomes that a given class of rearrangement operations can bring about.\cite{Hay2015, Hay2017, Kolmes2020Gardner, Kolmes2022Plateau}  
The second is to determine which class of rearrangement operations most appropriately captures the physics of a given physical system.\cite{Helander2017ii, Helander2020, Mackenbach2022, Mackenbach2023Measure, Mackenbach2023Miller, Kolmes2024Flutes}

If we can solve these two problems, then we can construct robust thermodynamic bounds on wave-particle interactions -- indeed, on any phase-space rearrangements -- for a variety of applications. 
These include efficiency limits for alpha channeling\cite{Fisch1992, Fisch1993, Fisch1995, Fetterman2008} and models for turbulent transport.\cite{Mackenbach2022, Mackenbach2023Measure, Mackenbach2023Miller} 
In fact, these rearrangement problems are closely connected with (and, in some cases, formally identical to) a variety of other problems both inside and outside plasma physics, in fields ranging from physical chemistry to the quantification of income inequality.\cite{Dalton1920, Horn1964, Atkinson1970, Berk1970, Bartholomew1971, Zylka1985, Morrison1989, Morrison1998, Thon2004, Aboudi2010, Lemou2012, Levy2014, Lostaglio2015, Brandao2015, Baldovin2016, Korzekwa2019, HoskingArXiv} 

This paper is based on an invited talk at the 2023 APS-DPP meeting, and very roughly follows the structure of that talk. 
Part of the paper functions as an introduction to the subject, summarizing advances over the last few years in understanding and applying theories of free energy in plasma systems. 
However, several parts of this paper have not been presented elsewhere. 
In particular, the explicit free-energy calculations in Section~\ref{sec:lossCones} are new; the application of free-energy theory to loss-cone modes was previously explored in Ref.~\onlinecite{Kolmes2024Flutes}, but that document was focused only on the thresholds at which the free energy vanishes. 
The proof of the maximum-energy ground state for the loss-cone-truncated Maxwellian in the same section is also new. 

This paper is structured as follows. 
Section~\ref{sec:definitions} defines the different free energies being considered. 
Section~\ref{sec:spectrum} discusses what is and is not known about the spectrum of ground states that can be accessed using diffusive exchange operations. 
Section~\ref{sec:lossCones} discusses how these free-energy theories play out in the context of loss-cone instabilities in a centrifugal mirror trap, and describes how to constrain the free energy in order to take into account the flute-like nature of many of these modes. 
Section~\ref{sec:discussion} is a discussion of the results. 

\section{Defining the Problem} \label{sec:definitions}

The key to quantifying the amount of energy that can be extracted from a system is to define the rules governing the ways in which that system may be rearranged. 
Different physical scenarios call for different rules for these rearrangements. 

\subsection{Two Basic Operations}

The earliest version of this theory for plasma physics was introduced by Gardner in 1963.\cite{Gardner1963} 
If the process underlying the rearrangement is Hamiltonian, then it must conserve the volumes of elements in phase space. 
Gardner's theory (sometimes called Gardner restacking) allows any rearrangement that conserves phase-space volumes. 
If the phase-space elements are rearranged such that the highest-population elements sit in the lowest-energy parts of phase space, then rearrangements of this kind can remove no further energy from the system. 
The energy that can be removed by transforming the initial state in this way is the Gardner free energy. 
The unique final state which has zero Gardner free energy is called the ground state. 

For a simple example, consider a discrete phase space consisting of three states, associated with energies $\varepsilon_0 = 0$, $\varepsilon_1 = 1$, and $\varepsilon_2 = 2$, respectively. 
If these states have initial populations $f_0 = 0$, $f_1 = 1$, and $f_2 = 2$, then a Gardner restacking procedure that transforms the system to its unique Gardner ground state is as follows: 
\begin{align*}
\begin{array}{|c|c|c|}
\hline 0 & 2 & 1 \\ \hline
\end{array} &\rightarrow 
\begin{array}{|c|c|c|}
\hline 2 & 0 & 1 \\ \hline 
\end{array}
\rightarrow 
\begin{array}{|c|c|c|}
\hline 2 & 1 & 0 \\ \hline 
\end{array} \, . 
\end{align*}
This maps the system from a state with energy $\sum_i f_i \varepsilon_i = 4$ to one with $\sum_i f_i \varepsilon_i = 1$, resulting in a release of 3 units of dimensionless energy. 

\begin{figure}
	\includegraphics[clip, trim=5cm 0pt 0pt 0pt, width=.8\linewidth]{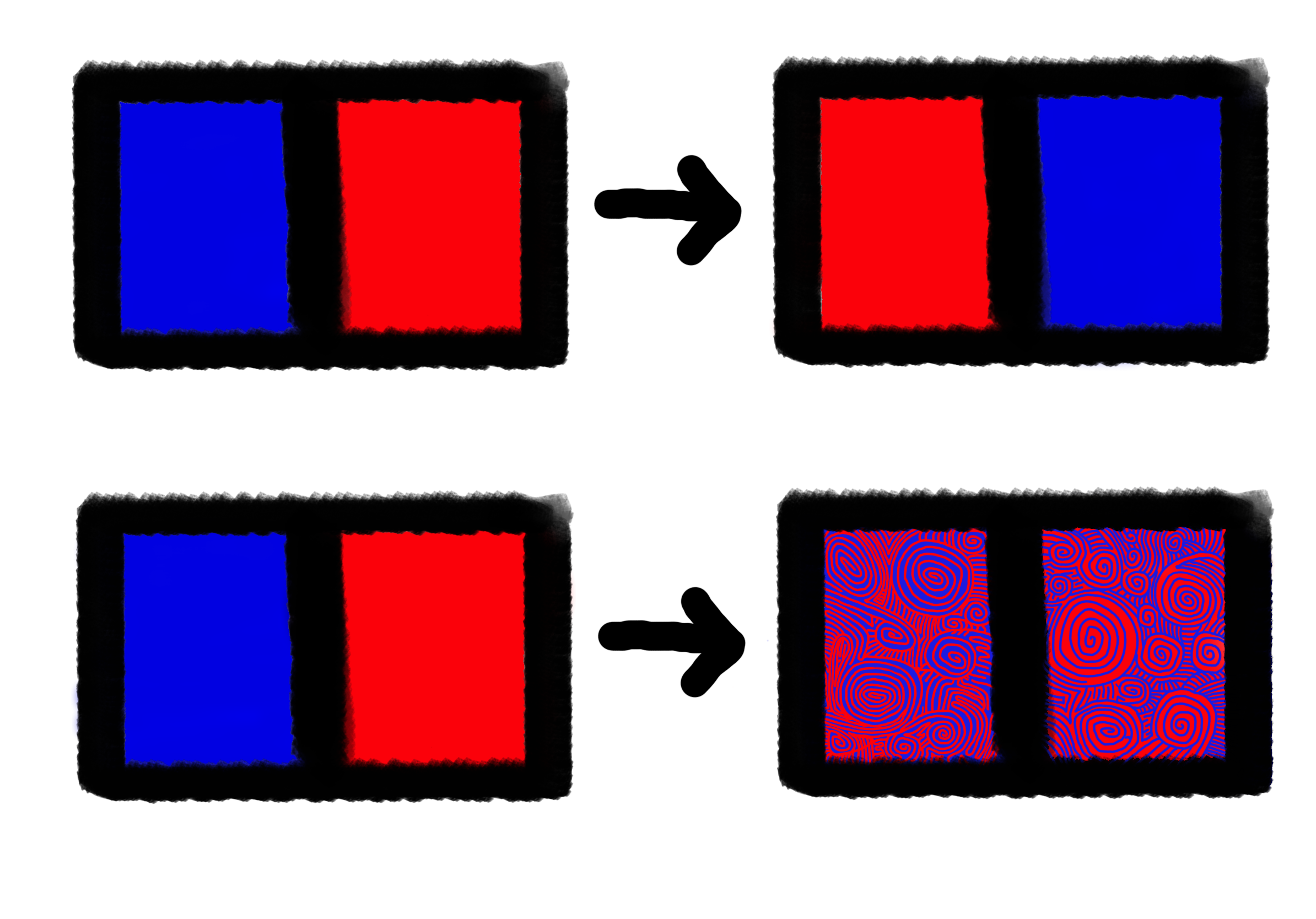}
	\caption{Above: Gardner's restacking operation consists of exchanging the populations in two elements of phase space. Below: the diffusive exchange operation consists of mixing the populations of two elements. Mixing of elements in phase space is commonly interpreted as the result of phase mixing on some small scale. }
	\label{fig:mixingCartoon}
\end{figure}

However, particularly for applications involving wave-particle interactions, the rearrangement of the distribution often involves phase mixing, wherein very fine-scale structures in phase space can lead to apparent smoothing effects. 
In other words, Liouville's theorem may still be respected on sufficiently microscopic scales, but on larger scales the dynamics can appear to be diffusive. 
This motivates an alternative rule, first proposed by Fisch and Rax,\cite{Fisch1993} in which phase space elements' contents are averaged rather than being exchanged. 
These two operations are illustrated in Figure~\ref{fig:mixingCartoon}. 

The maximum energy that can extracted from the same three-box system considered above can be accessed as follows: 
\begin{gather*}
\begin{array}{|c|c|c|}
\hline 0 & 2 & 1 \\ \hline 
\end{array}
\rightarrow
\begin{array}{|c|c|c|}
\hline 1/2 & 2 & 1/2 \\ \hline 
\end{array}
\rightarrow 
\begin{array}{|c|c|c|}
\hline 5/4 & 5/4 & 1/2 \\ \hline 
\end{array} \, . 
\end{gather*}
This releases $7/4$ units of energy rather than the $3$ that Gardner restacking operations were able to extract. 
Note, however, that we could have averaged the boxes in a different order, leading to a different ground state: 
\begin{gather*}
\begin{array}{|c|c|c|}
\hline 0 & 2 & 1 \\ \hline 
\end{array} 
\rightarrow 
\begin{array}{|c|c|c|}
\hline 1 & 1 & 1 \\ \hline 
\end{array} \, . 
\end{gather*}
This final state is clearly also a ground state, but it sits at a higher energy than the other, and corresponding to an energy release of only 1. 
Unlike Gardner restacking, diffusive mixing may lead a given initial state to any of a spectrum of possible ground states. 
Even for a simple three-box system like this one, there may be infinitely many such diffusively accessible ground states. 
The maximum possible energy release is called the diffusively accessible free energy. 

For a discrete system -- that is, a collection of $N$ boxes with populations -- the Gardner free energy is always greater than the diffusively accessible free energy. 
Hay, Schiff, and Fisch showed\cite{Hay2015, Hay2017} that although an $N$-box system may have an infinite number of diffuisvely accessible ground states, only a finite number of these could possibly correspond to the maximum energy release. 
In principle, then, one can simply write down the list of possible candidates and check which is best. 
However, the number of such candidates goes like $\mathcal{O}(N^{N^2})$. 
This makes a direct search very difficult for larger values of $N$. 

\subsection{Discrete and Continuous Phase Spaces}

There are some systems for which small $N$ is the case of greatest interest: for example, transitions between discrete atomic states.\cite{Fisch1993} 
However, for plasma physics applications, we are often interested in systems for which phase space is continuous. 
If we think of the discrete system as a coarse-grained approximation of a continuous system, this pushes us to the large-$N$ limit, where the approach of direct optimization is least tractable. 

It is also possible to define the continuous Gardner and diffusive-exchange problems directly in terms of a continuous system. 
The Gardner restacking problem can be formulated directly in terms of continuous curves,\cite{Dodin2005} and can be posed formally in terms of the ``symmetric decreasing rearrangement'' of the initial state.\cite{Riesz1930, HardyLittlewoodPolya, Brascamp1974, Almgren1989, Baernstein} 
One way of doing this is to describe the Gardner ground state $f_G$ as the decreasing function of energy alone that also satisfies 
\begin{gather}
\int \Theta [f_G(\varepsilon(\bx)) - \lambda] \, \D \bx = \int \Theta [f(\bx) - \lambda] \, \D \bx \quad \forall \lambda \in \mathbb{R}
\end{gather}
for the initial distribution $f$. Here $\bx$ is the phase-space coordinate (which generally includes position and velocity) and $\varepsilon(\bx)$ is the energy of a particle at $\bx$. 

The problem of maximum energy extraction under diffusive exchange, as it was originally posed by Fisch and Rax,\cite{Fisch1993} is to minimize 
\begin{gather}
W_\text{final} \doteq \lim_{t \rightarrow \infty} \int \varepsilon(\bx) f(\bx,t) \, \D \bx, 
\end{gather}
where $f(\bx, t)$ is the distribution at time $t$ and phase-space coordinate $\bx$, and where $f$ evolves according to 
\begin{gather}
\frac{\partial f}{\partial t} = \int K(\bx, \bx', t) \big[ f(\bx', t) - f(\bx, t) \big] \D \bx', 
\end{gather}
and the optimization is over all kernels $K(\bx, \bx', t)$ that are nonnegative and symmetric with respect to exchange of $\bx$ and $\bx'$. 
(Strictly speaking, the original formulation was in terms of a one-dimensional phase space in which $\bx$ is a scalar velocity $v$, but the generalization is straightforward). 

Although the original formulation is in terms of this space of kernels $K$, the space of possible kernels is large and in practice a direct search to find the optimal $K$ is prohibitively difficult. 
When trying to determine the optimal kernel $K$ -- whether to minimize the final energy, or with respect to any other metric -- it is typically advantageous to seek an indirect approach. 
This will be discussed in greater detail in Section~\ref{sec:spectrum}. 

\subsection{Additional Constraints}

Of course, real systems often obey a variety of constraints beyond, for example, phase-space volume conservation. 
For either of the two rearrangement operations discussed here, it is possible to impose additional constraints -- for example, by limiting which phase-space elements can be restacked or mixed with which. 
This idea was first introduced in the context of Gardner restacking by Helander;\cite{Helander2017ii, Helander2020} it works in essentially the same way for the diffusive exchange problem.\cite{Kolmes2020ConstrainedDiffusion}
Enforcing conservation of a given quantity for each particle means that exchanges are only allowed between pairs of elements with the same value of that quantity. 
This reduces the rearrangement problem to a set of independent rearrangement problems, each performed on the hyperplane of phase space on which the given quantity is constant. 
The inclusion of conservation laws for the appropriate adiabatic invariants turns out to be important in the application of the Gardner free energy for certain applications involving turbulence in magnetic confinement systems.\cite{Mackenbach2022, Mackenbach2023Measure, Mackenbach2023Miller} 
Constraints other than per-particle conservation laws are also possible;\cite{Kolmes2024Flutes} this will be discussed in more detail in Section~\ref{sec:lossCones}. 

\section{Characterizing the Spectrum of Diffusively Accessible Ground States} \label{sec:spectrum}

Originally, the diffusive-exchange problem was proposed in the context of alpha channeling, where waves are injected in order to remove energy from fusion products.\cite{Fisch1992, Fisch1993} 
For this reason, the early work on the subject largely focused on determining the upper bound on energy extraction.
This upper bound is what we mean when we refer to \textit{the} diffusively accessible free energy. 

This upper-bound problem (corresponding to the lower bound for the energy of the final ground state) was solved in Ref.~\onlinecite{Kolmes2020Gardner} for continuous systems. 
It turns out that in a continuous system, it is possible to construct a series of mixing operations that approaches the Gardner ground state arbitrarily closely. 
It is possible to show (and intuitively straightforward to see) that it is not possible to reach a lower-energy state than the Gardner ground state through mixing. 
Therefore, the Gardner free energy and the diffusively accessible free energy are identical in the continuous case. 

However, there is increasing interest in the application of free-energy calculations to phenomena like turbulence. 
For these applications, it is desirable not only to understand the largest possible energy release, but also to understand the full range of possible outcomes that can be brought about by mixing operations. 
This motivates the identification of the minimum stabilizing energy release, which is the lower bound on the possible release of energy that maps the initial state to a ground state. 
This can be posed equivalently in terms of the highest-energy accessible ground state. 
Note that although the original formulation of the problem allows mixing operations that can either increase or decrease the energy of the system (and it can be shown the mixing operations that increase the system energy are never necessary in order to reach the maximum possible release of energy\cite{Hay2015}), this minimum-energy-release problem is only interesting if ``annealing operations'' (those that deposit energy into the system) are prohibited. 
Annealing operations do not correspond to the expected behavior of natural modes, and if they are permitted it is typically possible to reach ground states with much higher energy than the initial state (infinitely higher, for most continuous systems, since one can simply mix the populated regions of phase space with empty regions at arbitrarily high velocity). 

The minimum stabilizing energy release is not known for arbitrary initial conditions, but it is known in certain particular cases. 
For the case of a bump-on-tail distribution, the minimum stabilizing energy release corresponds precisely to the classical quasilinear plateau solution, in which the region of the distribution with the population inversion is simply flattened.\cite{Kolmes2022Plateau} 
Similar plateau-like solutions can be shown to be optimal for certain close relatives of the bump-on-tail distribution. 
The minimum stabilizing energy release for loss-cone distributions will be discussed in Section~\ref{sec:lossCones}. 
For a discrete system, it is possible to enumerate the solutions to this problem when the system is small. 
This was done explicitly for the three-box system in Ref.~\onlinecite{Kolmes2022Plateau}. 

One corollary of the proof in Ref.~\onlinecite{Kolmes2020Gardner} is that any state that can be reached through Gardner restacking -- not just the Gardner ground state -- is also accessible (or arbitrarily close to being accessible) through diffusive exchange operations. 
It follows that any weighted average of these restacked states is also diffusively accessible. 
The Lynden-Bell equilibrium appears in statistical descriptions of astrophysical systems,\cite{LyndenBell1967, Ewart2022, Ewart2023} and can be understood as an average over an ensemble of systems that individually satisfy Liouville's theorem for some initial condition. 
Therefore, it also follows that the Lynden-Bell equilibrium is itself diffusively accessible. 

\section{Case Study: Flute-Like Loss-Cone Modes in Rotating Mirrors} \label{sec:lossCones}

The stabilization of flute-like loss-cone modes in rotating mirror configurations provides an interesting case study for how the theory of free energy may be applied in practice. 
There are two reasons for this. 
First, the intuition behind loss-cone modes revolves around rearrangement operations. 
The existence of an empty loss region alongside populated regions of phase space at higher energies means that it is possible to release energy by dropping particles from higher-energy trapped regions into lower-energy loss regions. 

Second, this system illustrates the role of constraints in the free-energy theory. 
By first computing the stabilization thresholds directly from the dispersion relations, then calculating the dependences of the free energy in these systems, it is possible to see whether or not the suppression of the free energy closely corresponds to the stability thresholds. 
In other words, it is possible to check how closely the suppression of the free energy corresponds to the stabilization of the various modes. 
As we will see, the basic form of the Gardner free energy does very poorly at explaining the stabilization thresholds of these modes. 
However, the inclusion of an additional constraint, taking into account the flute-like nature of the modes, leads to a much better-performing theory. 

\subsection{Modeling the Effects of Rotation}

To perform this calculation, it is necessary to write down a model for how the distribution function $f(\bv)$ depends on the mirror parameters. 
Consider a mirror-type configuration in which the magnetic field strength $B$ is maximized at the midplane and minimized at the edge, with the ratio of the maximum field strength to the minimum strength given by the mirror ratio $R$. 
Suppose the mirror is rotating, that is, suppose that a largely radial electric field crossed with the largely axial magnetic field causes the plasma to undergo drift in the azimuthal direction. 
Let $\Delta \Phi$ denote the difference in the combined centrifugal and electrostatic potentials along a field line between the edge and the midplane. 
Then, so long as the rotation frequency is small compared to the ion cyclotron frequency (so that corrections to the adiabatic invariants can be neglected),\cite{Volosov1969, Thyagaraja2009} the condition for a particle to be trapped can be written as 
\begin{gather}
(R-1) v_\perp^2 - v_{||}^2 + \frac{2 \Delta \Phi}{m} \geq 0 , \label{eqn:lossCone}
\end{gather}
where $v_{||}$ and $v_\perp$ are the velocity components parallel and perpendicular to the magnetic field and $m$ is the particle mass. 
There is no unique mapping between $(R, \Delta \Phi)$ and $f(\bv)$; the details of the distribution will depend on the particle sources, heating terms, and so on. 
However, it is sensible to expect $f(\bv)$ to vanish in regions of phase space where Eq.~(\ref{eqn:lossCone}) is not satisfied. 

A few models have been used for this problem in the literature.\cite{Volosov1969, Turikov1973, Kolmes2024Flutes} 
One that is both reasonably simple and matches Fokker-Planck simulations reasonably well for some choices of source\cite{Kolmes2024Flutes} is a Maxwellian, truncated so as to vanish inside the loss region: 
\begin{gather}
f_T(\bv) = A e^{-m v^2 / 2 T} \Theta \bigg[ (R-1) v_\perp^2 - v_{||}^2 + \frac{2 \Delta \Phi}{m} \bigg] . \label{eqn:truncatedMaxwellian}
\end{gather}
Here $A$ is a normalization constant (which depends on $R$ and $\Delta \Phi$), $\Theta$ is the Heaviside step function, and $T$ is the temperature. 
Spatial variations in $f$ are neglected, and the magnetic field is taken to be a square well, so that it does not vary in the interior of the mirror. 
For the sake of simplicity, we will focus on this model in this paper. 
For further discussion of alternatives, advantages, disadvantages, and numerical validation of the model, see Ref.~\onlinecite{Kolmes2024Flutes}. 
Note that the choice of model for $f$ is an important part of this calculation, and can have a significant impact on the resulting stability thresholds. 
It can be understood as a kind of initial condition for the analysis. 

It is often convenient to work with dimensionless coordinates. 
Let 
\begin{gather}
\bu \doteq \bv \sqrt{ \frac{m}{2 T} }
\end{gather}
and 
\begin{gather}
\phi \doteq \frac{\Delta \Phi}{T} \, . 
\end{gather}
Then $f_T$ can be rewritten as 
\begin{gather}
f_T(\bu) = A e^{-u^2} \Theta \big[ (R-1) u_\perp^2 - u_{||}^2 + \phi \big] .  
\end{gather}
Let $R > 1$ and $\phi \geq 0$, as these are the cases of greatest practical interest. 
Then if $f_T$ is normalized to $N$, 
\begin{align}
&A = N \bigg( \frac{m}{2 \pi T} \bigg)^{3/2} \nonumber \\
&\hspace{20 pt}\times \bigg\{ 
\text{erf}\sqrt{\phi}  - 2 \sqrt{\frac{\phi}{\pi}} e^{-\phi} \nonumber \\
&\hspace{50pt}+ 2 \sqrt{\frac{R-1}{\pi R}} e^{\phi/(R-1)} \Gamma\bigg( \frac{3}{2} , \frac{R \phi}{R-1} \bigg) \bigg\}^{-1} , 
\end{align}
where $\Gamma(a,b)$ is the incomplete gamma function. 

\subsection{Unconstrained Bounds} \label{sec:baseAE}

To begin, it is instructive to consider the Gardner free energy theory without any further modification. 
Largely following the notation in Ref.~\onlinecite{Helander2017ii}, define the level-set volume function 
\begin{gather}
H(\lambda) \doteq \int \Theta \big[ f(\bu) - \lambda \big] \D^3 \bu . 
\end{gather}
Within the trapped region, $f_T(u_\lambda) = \lambda$ when 
\begin{gather}
u_\lambda = \sqrt{ - \log \bigg( \frac{\lambda}{A} \bigg) } . 
\end{gather}
Then using spherical $(u, \theta, \varphi)$ coordinates in velocity space, we can calculate the level-set function for $f_T$ as follows: 
\begin{align}
&H(\lambda) = \nonumber \\
&\hspace{15pt}2 \pi \int_0^{u_\lambda} u^2 \, \D u \int_0^\pi \Theta \bigg[ R \sin^2 \theta - 1 + \frac{\phi}{u^2} \bigg] \sin \theta \, \D \theta . 
\end{align}
Define 
\begin{gather}
\lambda_* \doteq A e^{-\phi} . 
\end{gather}
Then 
\begin{align}
H(\lambda > \lambda_*) = \frac{4 \pi}{3} \bigg[ - \log \bigg( \frac{\lambda}{A} \bigg) \bigg]^{3/2}
\end{align}
and 
\begin{align}
&H(\lambda \leq \lambda_*) = \frac{4 \pi}{3} \phi^{3/2} \nonumber \\
&\hspace{10 pt}+ \frac{4 \pi}{3} \frac{\big[ - (R-1) \log(\lambda/A) + \phi \big]^{3/2} - (R \phi)^{3/2}}{\sqrt{R}(R-1)} \, . 
\end{align}
The Gardner ground state $f_G$ can be computed by setting 
\begin{gather}
H(f_G) = \frac{4 \pi}{3} \, u^3 . 
\end{gather}
This leads to 
\begin{widetext}
\begin{align}
f_G(\bu) = 
\begin{cases}
A e^{-u^2} & u < \sqrt{\phi} \\	
A \exp \bigg\{ \frac{1}{R-1} \bigg[ -  \big[ \sqrt{R}(R-1) \big( u^3 - \phi^{3/2} \big) +  (R \phi)^{3/2} \big]^{2/3} + \phi \bigg] \bigg\}  & u \geq \sqrt{\phi} \, . 
\end{cases}  \label{eqn:fG}
\end{align}
\end{widetext}
In order to calculate the energy released when $f_T$ is transformed to $f_G$, it is necessary to find the kinetic energy in each distribution. 

The energy $W_T$ in the initial distribution can be found analytically. 
\begin{align}
W_T &\doteq T \bigg( \frac{2 T}{m} \bigg)^{3/2} \int \D^3 \bu \, u^2 f_T . 
\end{align}
This can be evaluated to get 
\begin{align}
&W_T = \frac{3 AT}{2} \bigg( \frac{2 \pi T}{m} \bigg)^{3/2} \nonumber \\
&\times \bigg[ \text{erf}\sqrt{\phi} + \frac{R - 1 -(2/3)\phi}{\sqrt{R(R-1)}} e^{\phi/(R-1)} \text{erfc} \sqrt{\frac{R\phi}{R-1}} \, \bigg] , \label{eqn:WT}
\end{align}
where $\text{erf}$ is the error function $\text{erfc}$ is the complementary error function. 

In the limit of $\phi \rightarrow 0$, this is 
\begin{gather}
W_T \big|_{\phi = 0} = \frac{3 N T}{2} \, . 
\end{gather}
In this same limit, 
\begin{gather}
f_G = A \exp \bigg[ -\bigg( \frac{R}{R-1} \bigg)^{1/3} u^2 \bigg], 
\end{gather}
that is, the Gardner-restacked distribution is a Maxwellian with temperature $[R/(R-1)]^{1/3} T$, and the Gardner ground state has energy 
\begin{gather}
W_G \big|_{\phi=0} = \frac{3 N T}{2} \bigg( \frac{R}{R-1} \bigg)^{1/3}. 
\end{gather}
The resulting energy fractional energy release is 
\begin{gather}
\frac{W_T-W_G}{W_T} \bigg|_{\phi=0} = 1 - \bigg( \frac{R}{R-1} \bigg)^{1/3} . 
\end{gather}
Increasing $R$ reduces the available energy, as it narrows the loss cone and reduces the volume of empty phase space into which particles can be moved. 

In the opposite limit of $\phi \rightarrow \infty$, the loss cone vanishes, $f_T$ becomes a Maxwellian with temperature $T$, $A \rightarrow (m / 2 \pi T)^{3/2} N$, and $W_T \rightarrow 3 N T / 2$. 
In this limit, $f_G$ also becomes a Maxwellian with temperature $T$, so the Gardner free energy vanishes. 

\begin{figure*}
\centering
\includegraphics[width=.48\linewidth]{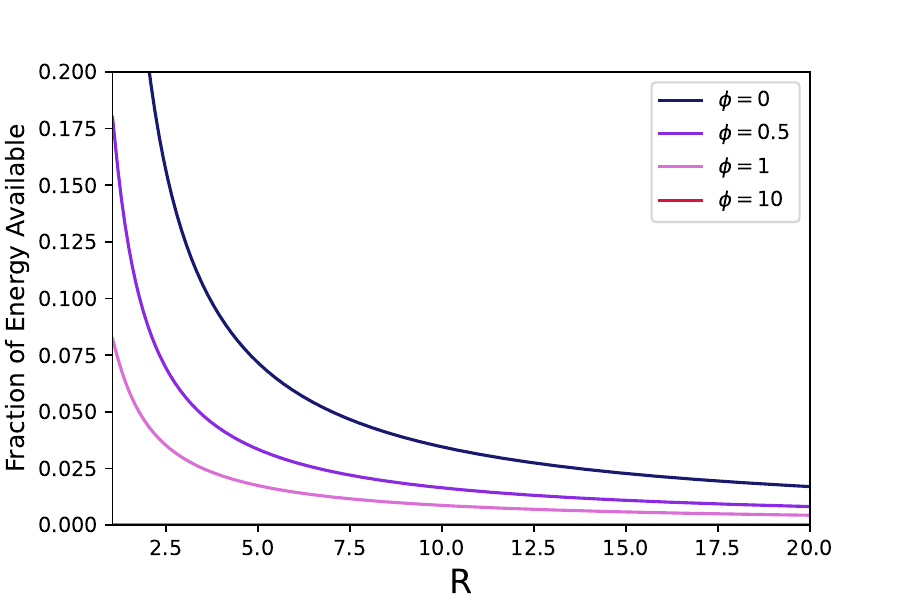} 
\includegraphics[width=.48\linewidth]{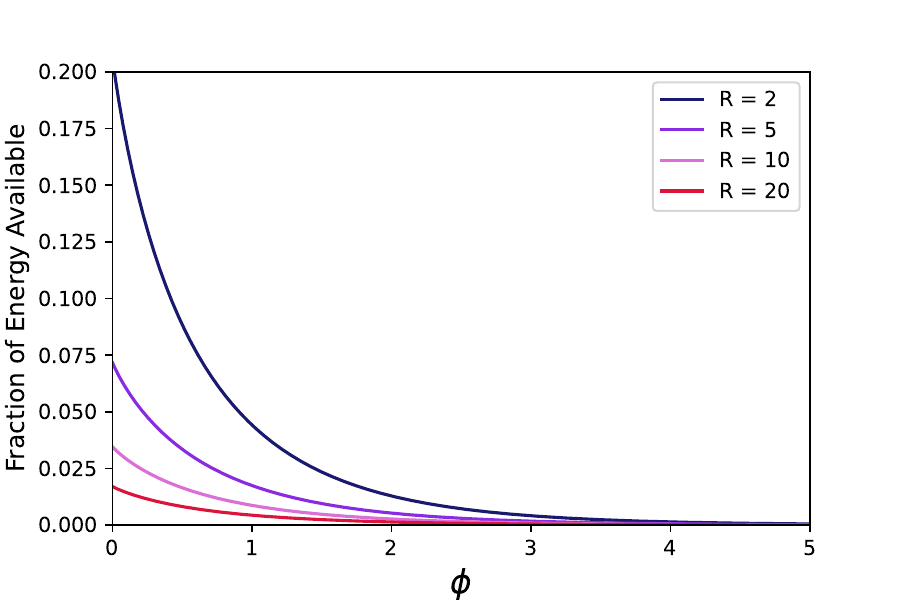}
\caption{The fraction of the total (initial) energy of the system that is extracted when the initial state is restacked to the Gardner ground state, as a function of the mirror ratio $R$ and the confining potential $\phi$, if no further constraints on the rearrangement are imposed. Larger $R$ and $\phi$ both suppress the free energy fraction. }
\label{fig:baseAE}
\end{figure*}

More generally, it is possible to calculate the kinetic energy in $f_G$ from Eq.~(\ref{eqn:fG}) numerically and compare the value with $W_T$ to get the fraction of the initial energy released. 
This is shown in Figure~\ref{fig:baseAE} for several choices of $R$ and $\phi$. 
The fraction of the initial energy that is released when the system is transformed to its Gardner ground state is a monotonically decreasing function of both $R$ and $\phi$. 

Recall from Section~\ref{sec:spectrum} that for a continuous system, the Gardner free energy also constitutes the least upper bound on the energy that can be released using diffusive mixing operations. 
However, we might also wish to know the minimum stabilizing energy release: that is, the least energy that can be released while still mapping $f_T$ to a ground state (or, more formally, the infimum of the accessible range). 
Interestingly, this bound is zero; it is possible to construct a sequence of mixing operations that results in a ground state while releasing vanishingly little energy. 

To see this, consider the part of the loss region with energy $\varepsilon$. 
Consider a series of mixing operations that mixes the contents of this shell with the part of the trapped region with energy $\varepsilon + \delta \varepsilon$, where $\delta \varepsilon$ is taken to be very small. 
It is possible to use mixing operations to equalize the density of $f$ throughout the part of the loss region with energy $\varepsilon$ and the part of the trapped region with energy $\varepsilon + \delta \varepsilon$, and to do so while releasing energy at every step. 
If every part of the loss region with energy $\varepsilon$ were homogenized with the part of the trapped region with energy $\varepsilon + \delta \varepsilon$ in this way, then in the limit where $\delta \varepsilon \rightarrow 0$, we would map the initial state to a ground state at the same energy. 

\subsection{Constraint for Flute-Like Modes} \label{sec:fluteConstraint}

With these results in hand, we can return to the question of how to match the free-energy results with the behavior of the loss-cone instabilities. 
Many of the most important of these modes are flute-like: their wave-numbers vanish in the direction of the magnetic field $(k_{||} = 0)$. 
Physically, this is related to the large mobility of the electrons in the direction of the magnetic field.\cite{Post1987} 
Flute-like loss-cone modes include the high-frequency convective loss cone (HFCLC), drift-cyclotron loss cone (DCLC), and Dory-Guest-Harris (DGH) modes.\cite{Dory1965, Rosenbluth1965, Post1966, Post1987, Kotelnikov2017}

In the limit of sufficiently large $\phi$, these instabilities must be suppressed: at some point, the loss region has been lifted to such high energies that no particles are affected by it. 
The key question, for present purposes, is whether the suppression of the instabilities correlates closely with the suppression of the free energy. 
In other words: is the free energy a good way of predicting the behavior of these modes? 
This is interesting as a way of testing the free-energy theory. 
It is also interesting for the purposes of understanding this class of modes. 
After all, there are many different loss-cone modes, so it would be very convenient to evaluate stability thresholds using a single free-energy metric rather than needing to check each mode's dispersion relation separately. 

Calculations of the HFCLC rotational stabilization criteria can be found in Refs.~\onlinecite{Volosov1969, Turikov1973}. 
Calculations of the HFCLC, DCLC, and DGH stabilization criteria can be found in Ref.~\onlinecite{Kolmes2024Flutes}. 
The details of the behavior of these modes is not the focus of this paper, but we will outline the key results as relevant for the comparison with the free-energy theory. 
\begin{enumerate}
	\item All three modes become stable when $\phi$ is sufficiently large and positive. 
	\item For distributions modeled by $f_T$, stabilizing values of $\phi$ are typically $\phi \leq 1$. 
	\item For distributions modeled by $f_T$, the HFCLC and DCLC stabilizing value of $\phi$ is higher when $R$ is higher. This behavior is not a unique feature of $f_T$, and also appears in other analytic and numerical models. (The DGH is typically only unstable for $f_T$ when $\phi \leq 0$). 
	\item For any distribution $f(v_{||},v_\perp)$, all three modes are stable if the projection $\int f \, \D v_{||}$ is a monotonically decreasing function of $v_\perp$. This is a sufficient condition for stability, but not a necessary condition. \label{point:projection}
\end{enumerate}
These points are discussed in greater detail in Ref.~\onlinecite{Kolmes2024Flutes}. 

Point by point, are these characteristics successfully captured by the unmodified free-energy theory discussed in Section~\ref{sec:baseAE}? 
\begin{enumerate}
	\item Yes. We can see from Figure~\ref{fig:baseAE} that the available energy vanishes when $\phi$ is large. 
	\item No. When $\phi = 1$, the available energy curves shown in Figure~\ref{fig:baseAE} may only be reduced by a factor of $\sim~1/4$ relative to their values at $\phi = 0$. This suggests that the mode might saturate at a lower level, but it does not suggest that the mode should vanish altogether. 
	\item No. The unmodified free-energy theory predicts the opposite trend, with lower free energy when $R$ is larger. 
	\item No. The unmodified free-energy theory predicts that a system is in a stable ground state when $f$ is a monotonically decreasing function of energy, not when its projection is monotonic. 
\end{enumerate}
The unmodified version of the free-energy theory does rather poorly at predicting when these modes will be stable. 
This is because it is missing a key constraint.\cite{Kolmes2024Flutes}

A hint can be found in the appearance of the projected distribution in point (\ref{point:projection}). 
The quasilinear velocity-space diffusion operator can be written as\cite{KrallTrivelpiece} 
\begin{align}
\frac{\partial f}{\partial t} \bigg|_{\text{QL}} &= \frac{\partial}{\partial \bv} \cdot \mathsf{D} \cdot \frac{\partial}{\partial \bv} \, f , 
\end{align}
where 
\begin{gather}
\mathsf{D} \doteq D_0 \int \frac{\omega_i \mathcal{E}_{\mathbf{k}}}{(\mathbf{k} \cdot \bv - \omega_r)^2 + \omega_i^2} \frac{\mathbf{k} \mathbf{k}}{k^2} \, \D \mathbf{k} . 
\end{gather}
Here $D_0$ is a species-dependent constant, $\mathcal{E}_\mathbf{k}$ is the spectral energy density, $\omega_r$ and $\omega_i$ are the real and imaginary parts of the wave frequency, and $\mathbf{k}$ is the wavenumber. 
Recall that the modes under consideration are all flute-like. 
If $k_{||} = 0$, then the quasilinear diffusion operator does not drive velocity-space diffusion in the parallel direction, and the operator does not distinguish between different values of $v_{||}$. 

This motivates a constraint on the allowed rearrangement operations.\cite{Kolmes2024Flutes} 
If the mixing is driven by flute-like modes, then we should allow only mixing in the perpendicular direction, and we should require that any rearrangement acting on a point $(v_{||}, \bv_\perp)$ must act identically on all other points with the same $\bv_\perp$. 
Intuitively, this means that flute-like rearrangements act on the projection $\int f \, \D v_{||}$ and cannot access any free energy associated with population inversions in the parallel direction. 

\begin{figure*}
	\includegraphics[width=.48\linewidth]{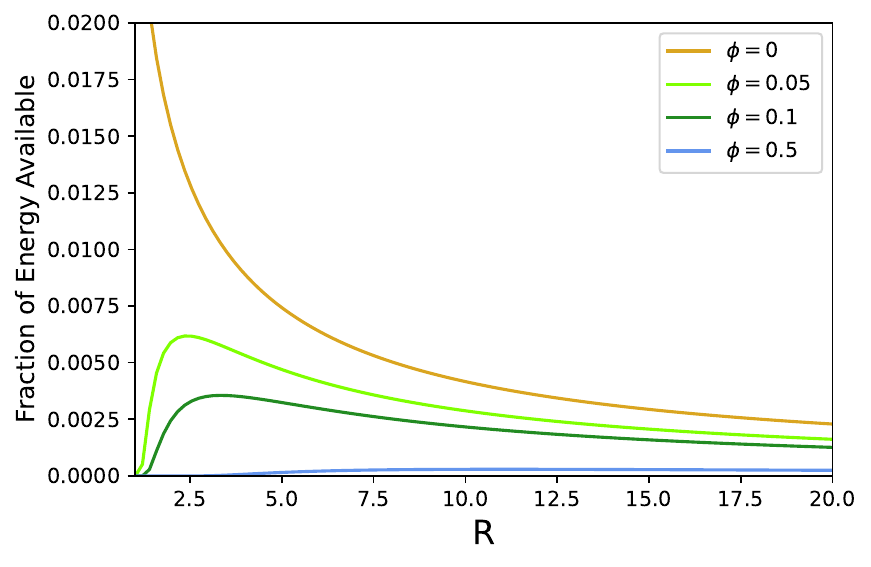}
	\includegraphics[width=.48\linewidth]{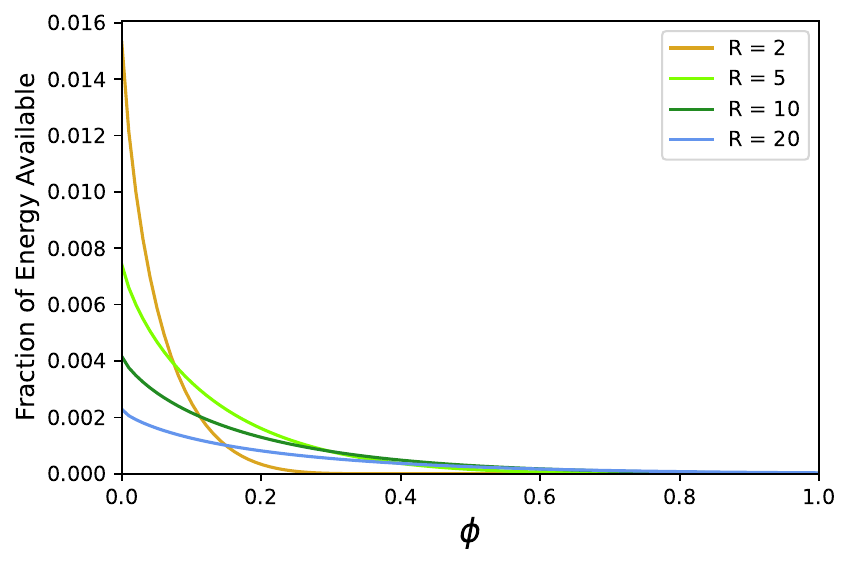} 
	\caption{The fraction of the total (initial) energy of the system that is extracted when the initial state is restacked to the Gardner ground state, subject to the additional ``flute-like" constraint that the rearrangements can only move phase space elements in the perpendicular direction and can only perform rearrangements that affect all values of $v_{||}$ for any given $v_\perp$. Larger $\phi$ always reduces the available energy, but the available energy may not be a monotonic function of $R$. } 
	\label{fig:fluteAE}
\end{figure*}

\begin{figure}
	\includegraphics[width=.96\linewidth]{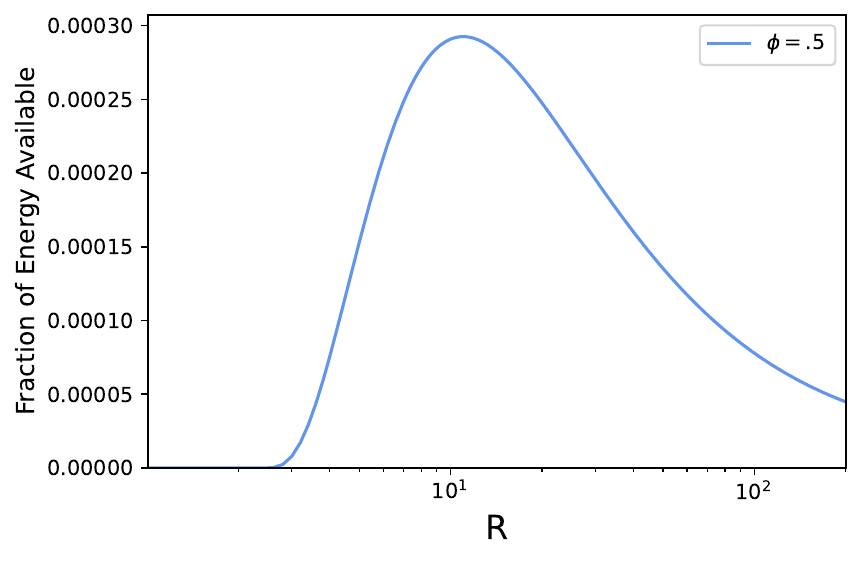}
	\caption{This is one of the particular cases shown in the left panel of Figure~\ref{fig:fluteAE}, with a magnified vertical axis and a logarithmic horizontal axis in order to highlight the non-monotonicity of the behavior more clearly. For this choice of $\phi$, the free energy entirely vanishes when $R$ is below some threshold. } 
	\label{fig:nonmonotonicity}
\end{figure}

Continuing to take $\phi \geq 0$ and $R > 1$, 
\begin{align}
\int_{-\infty}^{+\infty} f_T(\bu) \, \D u_{||} &= \sqrt{\pi} A e^{-u_\perp^2} \text{erf} \sqrt{(R-1) u_\perp^2 + \phi} \, . 
\end{align}
The projection is a monotonically decreasing function of $u_\perp$ if and only if 
\begin{gather}
\sqrt{\pi \phi} \, e^{\phi} \text{erf} \sqrt{\phi} \geq R-1. 
\end{gather}
The perpendicular energy that can be released by restacking operations with this ``flute-like'' constraint is equivalent to the original Gardner problem in two dimensions for the projected distribution $\int f_T \, \D u_{||}$. 
For the purposes of tracking the fraction of the total energy that is available, we also need to take into account the (entirely inaccessible) parallel component of the kinetic energy. 
This parallel component is 
\begin{align}
W_{T||} &\doteq T \bigg( \frac{2T}{m} \bigg)^{3/2} \int \D^3 \bu \, u_{||}^2 f_T \\
&= \frac{AT}{2} \bigg( \frac{2 \pi T}{m} \bigg)^{3/2} \bigg[ - \frac{2}{R} \sqrt{\frac{\phi}{\pi}} e^{-\phi} + \text{erf}\sqrt{\phi} \nonumber \\
&\hspace{30pt}+ \bigg( \frac{R-1}{R} \bigg)^{3/2} e^{\phi/R-1} \text{erfc}\sqrt{\frac{R\phi}{R-1}} \, \bigg] . 
\end{align}
Taking this into account, Figure~\ref{fig:fluteAE} shows the fraction of the energy that is available in $f_T$ under this new constraint. 
The constraint reduces the fraction of the energy that is available, but more importantly, it qualitatively changes the dependence of the free energy on the mirror ratio $R$. 
Before adding the constraint, the free energy was a monotonically decreasing function of $R$. 
After adding the constraint, it is generally non-monotonic, though it always becomes a decreasing function of $R$ when $R$ becomes sufficiently large. 
This is shown most clearly in the case highlighted in Figure~\ref{fig:nonmonotonicity}. 
The resulting stability condition was first introduced in Ref.~\onlinecite{Kolmes2024Flutes}. 

The non-monotonic dependence on $R$ is a surprise, but it is understandable. 
In fact, this non-monotonicity helps to resolve a discrepancy between the usual intuitions about the role of the mirror ratio and the actual behavior of the stabilization thresholds. 
If the free energy exists due to the loss cone, and larger $R$ means a smaller loss cone, then it would be reasonable to expect that larger $R$ should translate to less free energy. 
Indeed, this is what appears in Figure~\ref{fig:baseAE}, for the unconstrained free energy. 
But the $\phi$ thresholds for stabilizing flute-like loss-cone modes are \textit{higher} when $R$ is larger. 
The reason for this is that when $R$ decreases, there is more and more free energy, but less and less of it is accessible to the kinds of rearrangements that flute-like modes can produce. 
But the original intuition is recovered when we look at the large-$R$ limit: the modes may not stabilize, but the amount of free energy available to them eventually drops as $R$ increases (likely corresponding to modes that are unstable but saturate at a very low level). 
This happens because as $R$ increases, the loss-cone modes can access greater and greater fractions of the free energy, but there is less and less of it to begin with. 

\subsection{Modification to Account for Lost Particles}

\begin{figure}
	\centering
	\includegraphics[width=\linewidth]{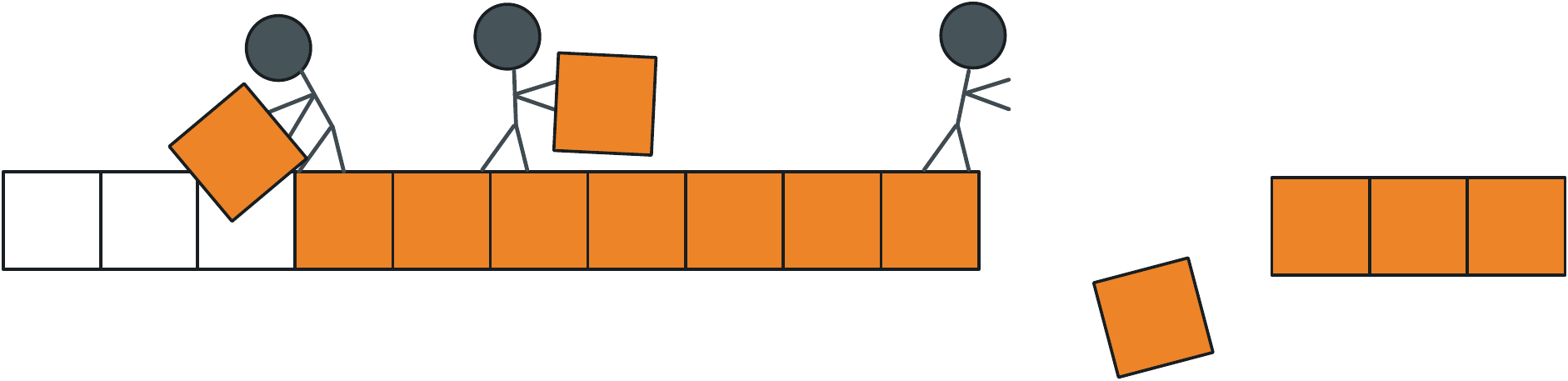}
	\caption{In a system with a loss cone, there are regions in phase space which do not ``fill up'' when phase space elements move into them.} \label{fig:chuckBox}
\end{figure}

The analysis thus far has treated loss-cone modes as being those modes which are associated with loss-cone distributions. 
In other words, the presence of a loss cone is associated with a particular class of kinetic distributions (those which vanish inside the loss cone), and distributions with this structure generally have population inversions which can drive instabilities. 
However, there is a meaningful distinction between the free energy in a system with an actual loss region and the free energy in a system for which the initial distribution simply happens to vanish in a given region. 
If a phase-space element is moved into a loss region, it promptly exits the system, and the loss region does not remain occupied. 
In the simplest case, it can be modeled as leaving the system without any additional interaction, so that whatever energy it has in its final phase-space position (inside the loss region) is carried away and not counted as available. 
This idea was first explored in Ref.~\onlinecite{Kolmes2024Flutes}, and is illustrated in Figure~\ref{fig:chuckBox}. 

In many cases, this can substantially increase the quantity of energy that can be extracted. 
For example, in a non-rotating mirror, the loss cone includes a region at vanishingly small energy, so in the absence of any additional constraints, Gardner restacking operations can release the entire energy content of the system. 

The effects of the loss-cone sink are perhaps most interesting in the case of the free energy with the ``flute-like" constraint discussed in Section~\ref{sec:fluteConstraint}. 
Without the loss region, the constraint prevents the system from performing rearrangements that distinguish between different values of $v_{||}$; entire columns of phase-space elements with a given $\bv_\perp$ must be moved together. 
With the loss region, it is possible to partially circumvent this constraint by moving the column of elements so that only part of it falls into the loss region. 
Then a segment of the column can be left behind while the surviving phase-space elements can be moved elsewhere. 

In other forms of the Gardner and diffusive-exchange theories, it is never necessary to perform so-called ``annealing" operations (that is, those that raise the energy of the system) in order to reach the minimum-energy ground state.\cite{Hay2015} 
However, when we include both a loss region and the flute-like constraint, there are scenarios in which the minimum-energy state can only be reached using sequences of rearrangements that include annealing (because these operations are sometimes necessary in order to drop off part of a column of phase space elements in the loss region). 

Interestingly, this implies that a configuration can be a ground state if the loss region is treated as unoccupied space, but have nonzero available energy if the loss region is instead treated as a sink. 
In other words, it suggests the existence of instabilities which rely on the loss of particles that get into the loss region, and it suggests that these instabilities would not be captured by an analysis which treated the loss region only as an initially unoccupied part of phase space (as, indeed, analytic treatments of loss-cone modes typically do). 

The appearance of these annealing operations motivates a distinction between strong and weak ground states. 
In a strong ground state, no sequence of rearrangements can possibly lead to a lower-energy state. 
In a weak ground state, no \textit{single} rearrangement operation can reduce the energy of the state. 
A state that would be a ground state in the absence of any loss-cone sink is always a weak ground state in the presence of the sink, but may not be a strong ground state. 
The stability conditions that appear in linear analyses of the HFCLC, DCLC, and DGH modes appear to more closely match the weak ground-state condition.\cite{Kolmes2024Flutes} 
This makes sense, given that these linear analyses did not include any particle sinks in the loss-cone region, instead treating the loss-cone as simply a region of phase space which happens to be unoccupied by the leading-order kinetic distribution. 

\section{Discussion} \label{sec:discussion}

Theories of free or available energy offer a way to calculate very general bounds on the possible behavior of plasma systems without the need to resolve (or even specify) all of the details of the dynamics of those systems. 
This can be useful in cases where the bound is computationally less expensive than the detailed calculation, as is often the case for applications involving turbulence.\cite{Mackenbach2022, Mackenbach2023Measure, Mackenbach2023Miller}
It can also be intrinsically useful to be able to draw conclusions about entire classes of systems with a single calculation. 
For example, there are many flute-like loss-cone modes,\cite{Post1987} and the calculations discussed in Section~\ref{sec:lossCones} should be applicable to all of them. 
The approach elaborated upon here, namely considering phase space rearrangements as a series of steps operating on a finite set of phase space volumes, has been shown to be particularly powerful despite its apparent simplicity. 
In the limit of many volumes, we were able to recover the surprising result that Gardner restacking could be achieved through diffusive operations. 
We were able also to recover theorems in a variety of constrained problems. 

For processes described by the diffusive-exchange operator, one major problem of interest is to characterize the spectrum of possible states that can be reached from a given initial condition.\cite{Fisch1993, Hay2015, Hay2017, Kolmes2020ConstrainedDiffusion, Kolmes2020Gardner, Kolmes2022Plateau} 
When phase space is continuous, the Gardner ground state is the lower bound for the final energy, and it is proved that it is possible to get arbitrarily close to that state diffusively.\cite{Kolmes2020Gardner} 
The highest-energy accessible ground state (when only energy-releasing operations are allowed) is known in certain cases but not others. 
Ref.~\onlinecite{Kolmes2022Plateau} shows that the quasilinear plateau solution is the highest-energy accessible ground state for the bump-on-tail distribution. 
This paper shows that the highest-energy ground state for a loss-cone-truncated Maxwellian is simply an isotropic Maxwellian at the same temperature. 

Another major problem of interest -- both for Gardner restacking and for diffusive exchange -- is to identify and understand the appropriate additional constraints to impose. 
On its own, the Gardner free energy is often a significant overestimate of how much energy is realistically extractable from a given system. 
One reason for this is that real systems often obey a variety of constraints in addition to Liouville's theorem. 
However, it is possible to formulate a constrained rearrangement problem in which other constraints are also considered. 
This was first done for the case of constraints that take the form of conservation laws applied to each phase-space element.\cite{Helander2017ii, Helander2020} 
However, there are also systems for which the relevant constraints take other forms. 

One example of this is the free energy associated with the loss region of a mirror configuration (rotating or otherwise). 
Many of the major loss-cone instabilities are flute-like, which means that they not only do not rearrange phase-space elements in the velocity direction parallel to the magnetic field, but they also cannot separately rearrange phase-space elements with different values of $v_{||}$. 
This leads to a more restrictive additional constraint on the allowed rearrangements. 
It turns out that this constraint greatly improves how well the Gardner free energy tracks the actual stability thresholds of the modes. 

Despite progress on these problems, there is work yet to be done. 
There remain open questions regarding the characterization of the spectrum of diffusively accessible states. 
In addition, there are many systems for which the question of which constraints are necessary (and how those constraints behave) has not yet been explored. 
As our understanding of these rearrangement processes improves, there are good reasons to hope that theories of free or available energy can be an increasingly practical tool with which to understand the behavior of plasma systems. 

\begin{acknowledgements}

The authors thank Robbie Ewart, Alex Glasser, Mike Mlodik, Ian Ochs, Jean-Marcel Rax, Tal Rubin, and Alex Schekochihin for helpful conversations. 
This work was supported by ARPA-E Grant No. DE-AR0001554. 
This work was also supported by the DOE Fusion Energy Sciences Postdoctoral Research Program, administered by the Oak Ridge Institute for Science and Education (ORISE) and managed by Oak Ridge Associated Universities (ORAU) under DOE Contract No. DE-SC0014664. 

\end{acknowledgements}

\section*{Data Availability Statement}

Data sharing is not applicable to this article as no new data were created or analyzed in this study.

\bibliographystyle{apsrev4-2} 
\bibliography{../../Master.bib}

\end{document}